\documentstyle[11pt]{article}
\title { Charmed baryons in bootstrap quark model }
\author{S.M.~Gerasyuta\thanks{Present address: Department of Physics,
 LTA, Institutski Per.5, St.Petersburg 194021, Russia }  ,
 D.V.~Ivanov \\
{\it Department of Theoretical Physics,}\\ {\it St. Petersburg
 State University}\\ {\it 198904, St.Petersburg, Russia }}
\date{}
\begin{document}

\maketitle

\begin{abstract}

 In the framework of dispersion relation technique the relativistic
 three-quark equations including heavy quarks are found. The approximate
 solutions of the relativistic three-particles equations based on the
 extraction of leading singularities of amplitudes are obtained. The
 mass values of S-wave multiplets of charmed baryons are calculated.

\end{abstract}

\section{Introduction}

 After detection of the $ J / \psi $-particles, the study of heavy quark
 properties   has
 become one of leading directions in particle physics . Some features of
 $ J / \psi $-meson and other states of charmonium stimulated the development
 of theory of strong interactions (QCD). The further progress is connected
 with detection of the bottomonium (family Y) and B-mesons (particles
 containing fifth b-quark).
 At present number of experimental and theoretical work is carried out,
 on the basis of which the uniform conception about dynamics of
 bound states of heavy quarks were obtained. The most
 reasonable and adequate approaches are the method of dispersion sum rules in
 QCD [1-5] and non-relativistic potential model [ 6-12 ], in the framework
 of which the quantitative description practically of all known
 properties of quarkonium $(Q\bar {Q})$ was received.

 Doubtless advantages of potential approach to description of heavy mesons
 are their simplicity and presentation. It permits to calculate ( in the
 framework of accepted model of potential ) position of levels, width of
 radiation transitions between levels and width of annihilation decays.
 Thus, one can find actually the behaviour
 of system in enough wide interval of distances, that is important for
 understanding of dynamics of quark interactions. The serious
 difficulty of potential approach is that it is impossible completely to
 be justified in the framework of QCD.

 The method of sum rules of QCD [1,2], based on complex non-perturbative
 structure of QCD vacuum was successfully used
 for the study of charmonium. To realize the spectrum  of quarkonium, it is
 necessary to know the quark masses and the nature of the connecting forces.
 According to the method of sum rules of QCD, quarks exist in the complex
 medium -- non-perturbative vacuum, which densely populated by
 long-wave fluctuations of gluon field. These non-perturbative fluctuations
 determine the effective quark attraction.
 The sum rules of QCD permit to calculate the
 effect of long-wave gluon fluctuations and to make exact predictions for
 lowest levels with a different quantum numbers in terms of
 fundamental parameters. The most considerable difficulty of application
 of the sum rules method to the charmonium is account of the
 relativistic corrections. It is necessary to take into account
 with purely relativistic effects ( described by Breit-Fermi
 hamiltonian) radiative corrections of higher orders. Such
 calculation is so far away.

 The problem of the charmed baryons is investigated considerably less. The
 sum rules of QCD were applied for the first time to the charmed baryons in
 the paper [13]. However consideration was carried out within the limits of
 infinite mass of c-quark and also the contribution of the gluon
 condensate was not taken into account in this case.
 The subsequent calculations have allowed
 to calculate the masses of the S-waves charmed baryons with one c-quark
 [14,15].The potential quark models enable to calculate the mass spectrum
 of the charmed baryons of multiplets $J^{P} = {\frac{1}{2}}^{+},
 {\frac{3}{2}}^{+} $ [16-26].
 The most considerable difficulty of
 application of these methods is account of relativistic corrections
 at consideration of system of heavy and light quarks.

 The present paper is devoted to calculation of mass spectrum of charmed
 baryons in the framework of the bootstrap quark model.
 The relativistic Faddeev equations are constructed
 in the form of the dispersion relations over the two-body subenergy.
 The approximate solution of integral three-quark equations are obtained by
 taking into account two-particle and triangle singularities, all the weaker
 ones being neglected. The mass values of charmed low-lying multiplets
$ J^P ={\frac{1}{2}}^{+}, {\frac{3}{2}}^{+} $ are calculated.

 In section 2 the relativistic three-particle equations,
 which describe the interaction of light and
 heavy quarks in the charmed baryons are obtained.
 Section 3 is devoted to discussion of calculation results of the low-lying
 charmed baryons mass spectrum.
 In Conclusion we discuss the status of the considered model.

 In Appendix~A  the three-particle integral equations for the
 charmed baryons of multiplets $J^{P} = {\frac{1}{2}}^{+}, {\frac{3}{2}}^{+}$
 are given.
 In Appendix~B we present the approximate algebraic equations for the charmed
 baryons, accounting only the contribution of two-particle and triangle
 singularities in the scattering amplitude.

\section { Three-particle quark amplitudes of the S-wave charmed
baryons}

 In the papers [27,28]  relativistic integral three-particle
 Faddeev equations was constructed in the form of the dispersion relations
 over the two-body subenergy. Using the method of extraction of the main
 singularities of amplitude the mass spectrum of S-wave baryons,
 including u, d, s-quarks is calculated. In present paper this method
 is applied for study of the mass spectrum of baryons, including c-quarks.

 For the clarity we derive the relativistic Faddeev equations
 for the particle ${\Sigma}^{+}_{c} (J^{P} =
 {\frac{1}{2}}^{+})$. We construct the amplitudes of three-quark
 interaction, which include two-particle interactions with the
 quantum numbers of diquarks $J^{P}=0^{+},1^{+}$.

 Let us consider the baryon current which create three quarks (Fig.1 a).
 Subsequent two-quark interactions lead to the diagrams, shown in Fig.1 b-f.
 These diagrams might be divided into three groups so in each group the same
 pair of particles undergoes the last interaction (Fig.2).
 Consideration of diagram in Fig.1 allows to obtain the equation for
 three-particle amplitudes $A_{J}(s, s_{ik})$, ($J^{P} =0^{+}, 1^{+}$)
 in the graphic form (Fig.3). Here the invariants $ s_{ik}={(p_i +
 p_k)}^2 $, $ s={(p_i + p_j + p_k)}^2 $ for the quarks with
 momenta $ p_i, p_j, p_k $ are given.

 In order to present the amplitude $ A_{J}(s, s_{ik}) $ in the form of the
 dispersion relation explicitly, it is necessary to define the amplitude
 of two-quark interaction $ a_J (s_{ik}) $. We use the results of bootstrap
 quark model [29] and write down the diquark amplitude $ a_J (s_{ik}) $
 in the following way:

\begin{equation}
a_{J} (s_{ik})=\frac{G^{2}_{J} (s_{ik})}{1-B_{ J} (s_{ik})},
\end{equation}

\begin{equation}
B_{J} (s_{ik})=\int_{{(m_i + m_k)}^2}^{\infty} \frac{d{s}
_{ik}^{\prime}}{\pi} \frac{{\rho}_{J} (s_{ik}^{\prime })
 G^{2}_{J} (s_{ik}^{\prime})}{s_{ik }^{\prime} -s_{ik}}
\end{equation}

$$ {\rho}_{J} (s_{ik})=\frac{{(m_i + m_k)}^2}{4\pi} [ \gamma
(J^P) \frac{s_{ik}}{{(m_i + m_k)}^2} + \beta (J^P) + \frac {
\delta (J^P)}{s_{ik}} ] \times $$

\begin{equation}
\times \frac{\sqrt{(s_{ik} - {(m_i + m_k)}^2) (s_{ik} - {(
m_i - m_k)}^2)}}{s_{ik}}
 \end{equation}

 Here $ G_J $ is the vertex function of diquark, $ B_{J} (s_{ik}) $ is
 the Chew-Mandelstam function [30], $ {\rho}_{J} (s_{ik}) $ is
 the phase space of diquark (see Table 1).

 In the case in question the interacting quarks do not produce the
 bound state, then the integration in the dispersion integrals
 (4)-(6) is carried out from $ {(m_1 + m_2)}^2 $ to $\infty $. To write
 the system of equations (4)-(6) (Fig.3), it is necessary to account the
 spin-flavour part of the charmed baryon $ {\Sigma}_{c}^{+}$ wave function:
 $ |{\Sigma}_{c}^{+} \rangle =\sqrt{\frac{2}{3}} |\{
 u\uparrow d\uparrow c\downarrow \} \rangle -\frac{1}{\sqrt{6}} | \{
 u\uparrow d\downarrow c\uparrow \} \rangle - \frac{1}{\sqrt{6}} | \{
 u\downarrow d\uparrow c\uparrow \} \rangle  $.
 As a result we receive the system of integral equations :

$$ A_{1}(s,s_{12})=\hat{\alpha} b_{1}(s_{12}) L_{1}(s_{12})+
K_{1}(s_{12}) [\frac{1}{4} A_{1}^{(c)}(s,s_{13}^{\prime})+
\frac{3}{4} A_{0}^{(c)}(s,s_{13}^{\prime})+  $$
\begin{equation}
+\frac{1}{4} A_{1}^{(c)}(s,s_{23}
^{\prime})+ \frac{3}{4} A_{0}^{(c)}(s,s_{23}^{\prime})] \end{equation}

$$ A_{1}^{(c)}(s,s_{13})=\hat{\alpha} b_{1}^{(c)}(s_{13}) L_{1}^{(c)}(s_{13})
+ K_{1}^{(c)} (s_{13}) [\frac{1}{2} A_{1}(s,s_{12}^
{\prime})+\frac{3}{4} A_{0}^{(c)}(s,s_{12}^{\prime})-  $$
\begin{equation}
- \frac{1}{4} A_{1}^{(c)}(s,s_{12}
^{\prime})+ \frac{1}{2} A_{1}(s,s_{23}^{\prime})+ \frac{3}{4} A_{0}^{(c)}
(s,s_{23}^{\prime})- \frac{1}{4} A_{1}^{(c)}(s,s_{23}^{\prime})]
\end{equation}

$$ A_{0}^{(c)}(s,s_{23})=\hat{\alpha} b_{0}^{(c)}(s_{23})
L_{0}^{(c)}(s_{23})+ K_{0}^{(c)} (s_{23})
[\frac{1}{2} A_{1}(s,s_{13}^{\prime})+
\frac{1}{4} A_{1}^{(c)}(s,s_{13}^{\prime})+ $$
\begin{equation}
+\frac{1}{4} A_{0}^{(c)}(s,s_{13}
^{\prime})+ \frac{1}{2} A_{1}(s,s_{12}^{\prime})+ \frac{1}{4} A_{1}^{(c)}
(s,s_{12}^{\prime})+ \frac{1}{4} A_{0}^{(c)}(s,s_{12}^{\prime})]
\end{equation}
\vspace{0.5cm}

\noindent
 In eq.(4)-(6) we introduce functions
\begin{equation}  L_{J}(s_{ik})=\frac{G_{J}(s_{ik})}{1-B_{J}(s_{ik})}
\end{equation}
 and integral operators   \begin{equation}
 K_{J}(s_{ik})=L_{J}(s_{ik}) \int_{{(m_i + m_k)}^2}
^{\infty} \frac{d s_{ik}^{\prime}}{\pi} \frac{{\rho}_{J}(s_{ik}^{\prime})
G_{J}(s_{ik}^{\prime})}{s_{ik}^{\prime}- s_{ik}} \int_{-1}^{1} \frac{dz}{2}
\end{equation}

\noindent
 Here {\large $ b_{J} (s_{ik})= \int_{{(m_i + m_k)}^2}^{\infty }
 \frac{d s_{ik}^{\prime}}
{\pi} \frac{{\rho}_{J} (s_{ik}^{\prime}) G_{J} (s_{ik}
^{\prime})}{s_{ik}^{\prime} -s_{ik}} $ } is truncated Chew-Mandelstam
 function, z is cosine of the angle between the relative momentum of the
 particles 1 and 2 in the intermediate state and that of the particle 3 in the
 final state, which is taken in the c.m. of the particles 1 and 2. The
 coefficient $ \hat{\alpha} $ corresponds to the vertex of three-quark
 creation (later its value is not used ). The index ''c'' shows the
 presence of one (c) or two (cc) charmed quarks in diquark.

 Invariants $ s_{13}^{\prime} $ and $ s_{12}^{\prime} $ for various
 quarks are connected by the relation :

$$ s_{13}^{\prime} = m_1^2 + m_3^2 -\frac{(s_{12}^{\prime} +
m_3^2 -s) (s_{12}^{\prime } + m_1^2 - m_2^2)}{2 s_{12}^{ \prime}} \pm $$
\begin{equation}        \pm \frac{z}{2 s_{12}^{\prime}}{[(s_{12}^{\prime
} - {(m_1 + m_2 )}^2) (s_{12}^{\prime} - {(m_1 - m_2 )}^2)]}^
{1/2} \times           \end{equation}
$$\times {[(s_{12}^{\prime} - {(\sqrt{s} + m_3 )}^2) (s_{12}^
{\prime} - {(\sqrt{s} - m_3 )}^2)]}^{1/2} $$

 The system of integral equations for other charmed baryons of the
 multiplets $ J^P = {\frac{1}{2}}^{+}, {\frac{3}{2}}^{+} $
 are given in Appendix~A.

 We extract the diquark singularity of the amplitude $ A_J (s, s_{ik}) $:

\begin{equation}
A_J (s, s_{ik})=\frac{{\alpha}_J (s, s_{ik}) b_J (s_{ik})
G_J (s_{ik})}{1- B_J (s_{ik})} \end{equation}

 Then, for example, first equation of system (4)-(6) is possible to be
 rewritten for the amplitude $ {\alpha}_{1} (s, s_{12}) $ in the form:

$$ { \alpha } _ { 1} ( s, s_ { 12 } )=\hat { \alpha } + \frac { 1} { b_1 ( s_
{ 12 } ) } \int_ {{ ( m_1 + m_2 ) }^2}^ {{ \Lambda } _1 ( 1,2 ) } \frac { d s_
{ 12}^ { \prime }} { \pi } \frac {{ \rho } _1 ( s_ { 12}^ { \prime } ) G_1 (
s_ { 12}^ { \prime } ) } { s_ { 12}^ { \prime } -s_ { 12 }} \times $$

\begin{equation}
\times \int_{-1}^{1} \frac{dz}{2} \left ( \frac{G_1^{(c)}(
s_{13}^{\prime}) b_1^{(c)} (s_{13}^{\prime})}{1-B_1^{(
c)} (s_{13}^{\prime})} \frac{1}{2}
{\alpha}_1^{(c)}(s,s_{13}^{\prime})+
 \frac {G_0^{(c)} (s_{13}^{\prime})
B_0^{(c)} (s_{13}^{\prime})}{1-B_0^{(c)} (s_{13}^{
\prime})} \frac{3}{2}{\alpha}_0^{(c)} (s, s_{13}^{\prime
})\right ) \end{equation}

 The formula for $ s_{23}^{\prime}$ is similar to (7) with $ z $
 replaced by $ -z $. Therefore in (4) it is possible to
 replace $ [ A_1^{(c)} (s, s_{13}^{\prime}) + A_1^{(c)} (s,
 s_{23}^{\prime})]$ by $2 A_1^{(c)} (s, s_{13}^{\prime})$.
 In equation (11) we introduce the cut-off parameter $ {\Lambda}_J (i, k)$ at
 large value of $ s_{12}^{\prime}$.
 We choose the ''hard'' cutting, but we can use also the ''soft'' cutting,
 for instance $ G_{1}(s_{12})= G_{1} exp(-{(s_{12}-4 m^2 )}^2 /{\Lambda}^{2}_
 {1}) $, and do not change essentially the calculated mass spectrum.
 Similarly two other equations of system (5), (6) for reduced amplitudes
 $ {\alpha}_1^{(c)} (s, s_{13})$, $ {\alpha}_0^{(c)} (s, s_{23})$ are obtained
 (Appendix~B).

 The construction of the approximate solution of the system of equations
 (4) - (6) is based on the extraction of the leading singularities which are
 close to the region $ s_{ik} = {( m_i + m_k)}^2 $.
 Amplitudes with different number of rescattering (Fig.1) have the following
 structure of singularities. The main singularities in $ s_{ik}$ arise
 from pair rescattering of the particles $i$ and $k$. First of all there are
 threshold square root singularities. Also possible a pole singularities which
 correspond to the bound states. Figs.1b,c have only such two-particle
 singularities. The diagrams in Fig.1d,e apart from
 two-particle singularities have their own specific triangle singularities.
 The diagram in Fig.1f describes still more three-particle singularities.
 Apart from the two-particle and triangle singularities it has its own
 singularity. It is weaker than singularities of the diagrams Fig.1b-e.

 Such classification allows us to search the
 approximate solution of equations (4) - (6) by taking into account some
 definite number of leading singularities and neglecting all the weaker ones.
 We consider the approximation, which corresponds to the single interaction
 of all three particles (two-particle and triangle singularities).

 The contours of integration in complex plane of variable $ s_{13}^{\prime}$
 are given in Fig.4. At $ {( m_1 + m_2)}^2 \leq s_{12}^{\prime}
 \leq {(\sqrt{s} + m_3)}^2 $ the integration performed over the contour 1,
 and at $ {(\sqrt{s} + m_3)}^2 \leq s_{12}^{\prime} \leq {\Lambda}_J (1,2) $
 over the contour 2.

 The function $ {\alpha}_{J}(s,s_{12}) $ is a smooth function of $ s_{12} $
 as compared with the singular part of the amplitude, hence it can be
 expanded in a series in the singularity point and only the first term of
 this series should be employed further.
 It is convenient to take the middle point of Dalitz-plot region which
 corresponds to $ z=0 $. Then one determines the function $ {\alpha}_J
 (s, s_{12})$ and truncated Chew-Mandelstam function $ b_J (s_{12}) $ at the
 point $ s_{12}=s_0 = (s + m_1^2 + m_2^2 + m_3^2) / (m_{12}^2 + m_{13}^2 + m_
 {23}^2)$, where $ m_{ik} =\frac{1}{2}{(m_i + m_k)}, i=1,2,3 $.
 Such a choice of point $ s_0 $ allows us to replace the system of integral
 equations (4)-(6) by the algebraic system of equations for $ {\Sigma}_c^{+}$:

\begin{equation}
{\alpha}_1 (s, s_0)=\hat {\alpha} + \frac{1}{2} I_{1,1} (s, s_0
) \frac{b_1^{(c)} (s_0)}{b_1 (s_0)}{\alpha}_{1}^{(c)
} (s, s_0) + \frac{3}{2} I_{1,0} (s, s_0) \frac{b_0^{(c)}(
s_0)}{b_1 (s_0)}{\alpha}_0^{(c)} (s, s_0)  \end{equation}

$$ {\alpha}_1^{(c)} (s, s_0)=\hat{\alpha} + I_{1,1}^{(c)}(
s, s_0) \frac{b_1 (s_0)}{b_1^{(c)} (s_0)}{\alpha}_1 (s,s_0)-
$$ \begin {equation}  -\frac{1}{2} I_{1,1}^{(c)} (s, s_0)
{\alpha}_1^{(c)} (s, s_0) + \frac{3}{2} I_{1,0}^{(c)}(
s, s_0) \frac{b_0^{(c)} (s_0)}{b_1^{(c)} (s_0)}{ \alpha
}_0^{(c)} (s, s_0) \end{equation}

$$ {\alpha}_0^{(c)} (s, s_0)=\hat{\alpha} + I_{0,1}^{(c)}(
s, s_0) \frac{b_1 (s_0)}{b_0^{(c)} (s_0)}{\alpha}_1 (s,s_0) +
$$ \begin{equation} + \frac{1}{2} I_{0,1}^{(c)} (s,
s_0) \frac{b_1^{(c)} (s_0)}{b_0^{(c)} (s_0)}{\alpha}_1^
{(c)} (s, s_0) + \frac{1}{2} I_{0,0}^{(c)} (s, s_0){
\alpha}_0^{(c)} (s, s_0) \end{equation}

 Here
 \begin{equation} I_{J_1,J_2} (s, s_0)=\int_{{( m_i + m_k )
}^2}^{{\Lambda}_J (i, k)} \frac{d s_{ik}^{\prime}}{\pi}
\frac{{\rho}_{J_1} (s_{ik}^{\prime})}{s_{ik}^{\prime
} -s_0 } G_{J_1} \int_{-1}^{1} \frac{dz}{2} \frac{G_{J_2}}{
1-B_{J_2} (s_{ij}^{\prime})} \end{equation}

 In our approximation vertex functions $ G_J (s_{ik})$ are the constants.

 The function $ I_{J_1, J_2} $ takes into account singularities which
 corresponds to the simultaneous vanishing of all propagators in the
 triangle diagrams like those in Fig.1d,e. The solution of the system of
 equations (12) - (14) permits to determine the pole in $s$ (zero of
 determinant) which corresponds to the bound state of the three quarks.

 Similarly to the case of $ {\Sigma}_{c}^{+}$--baryon it is
 possible to obtain the rescattering amplitudes for all S-wave charmed
 baryons with $J^P = {\frac{1}{2}}^{+}, {\frac{3}{2}}^{+}$
 including  u, d, s, c-quarks, which satisfy the systems of integral
 equations (Appendix~A). If we consider the approximation, which corresponds
 to taking into account two-particle and triangle singularities, and define
 all the functions of the subenergy variables in the middle point of the
 physical region of Dalitz-plot $ s_0 $ then the problem reduces to one of
 solving simple algebraic system equations (Appendix~B).

 In the calculation in question the quark masses $ m_u =m_d =m, m_s, m_c $
 are not fixed. We fix such quark mass values, which allows us to calculate
 the masses of charmed baryons using the vertex constants and the cut-off
 parameters:

\begin{equation}
g_{J}^{(ik)}=\frac{m_i + m_k}{2\pi} G_{J}^{(ik)},{
\lambda}_{J}^{(ik)}=\frac{4{\Lambda}_J (i, k)}{{(m_i +
m_k)}^2}, s_0 =\frac{4 s_{ik}}{{(m_i + m_k)}^2},
\end{equation}

\noindent  where $J=0,1$; $ m_i$ and $ m_k $ are quark masses in the
 intermediate state of the corresponding quark loop.

\section {Calculation results}

 The masses of S-wave charmed baryons $ (J^P = {\frac{1}{2}}^{+},
 {\frac{3}{2}}^{+})$ including light u, d, s and heavy c-quarks
 are obtained. Two dimensionless cut-off parameters
 $ {\lambda}_q =10.7$ and $ {\lambda}_c =6.5$ are chosen for light quarks and
 c-quark respectively .
 For the diquarks, which include one light and one heavy quark, the cut-off
 parameter is equal $ {\lambda}_{qc} =\frac{1}{4}{(\sqrt{{\lambda}
_c} + \sqrt{{\lambda}_q})}^2 $. For light quarks the calculated values of
 vertex constants $g_1 =0.56 $ and $g_0 =0.70$ have coincided with ones,
 obtained earlier in the bootstrap quark model for
 light mesons and baryons [27-29]. This coincidence is related with the
 main contribution of one-gluon exchange in the vertex functions
 and is determined by the rules of $1/N_c $-expansion [31,32].
 For diquarks containing c-quark one common constant $g_c =0.857 $ is
 obtained. The model parameters are determined by experimental
 values of masses $ {\Lambda}_c, {\Sigma}_c, {\Xi}_{c}^{
 (A)}, {\Sigma}_{c}^{*}, {\Xi}_{c}^{*} $ [33]. The quark masses
 $ m_u = m_d =m=$0.495~GeV, $m_s =$0.77~GeV and $ m_c =$1.655~GeV
 determine the region, in which it is possible to calculate the mass
 spectrum of all S-wave charmed baryons.

 The results of calculations are shown in Table 2. In the parentheses are
 mentioned the experimental data [33]. In the calculations we assume
 the absence of the mixing of $ {\Xi}_{c}^{(A)}$ and $ {\Xi}_{c}^{(S)}$
 baryons, that is in agree with result of the paper [34].
 The fact is interesting, that the behaviour of Chew-Mandelstam functions
 which determine the dynamics of quark interactions for particles $ {\Xi}_{c}
 ^{(S)}$, is considerably different from one for other S-wave charmed baryons.
 This state is excited [18] and at considered values of parameters do not
 arise a bound state. In order to obtain the bound excited state we introduce
 the quark mass shifts, which take into account the confinement potential
 effectively [35].

 The change of quark mass values $ m_{u}^{*}=$0.41~GeV, $m_{s}^{*}=$0.985~GeV,
 $m_{c}^{*}=$1.4~GeV lead to the change of behaviour of
 Chew-Mandelstam functions and allows to receive the bound state
 for $ {\Xi}_{c}^{(S)}$-baryon.

 In considered calculation the masses of charmed baryons, containing
 two s-quarks, are overestimated in comparison with the papers [14-18].
 It happens according to the choice of the large value of s-quark mass,
 which allows to obtain the correct splitting of states $ {\Sigma}_c $
 and $ {\Xi}_{c}^{(A)}$.

\section{Conclusion}

 In the present paper the suggested method of the approximate solution of the
 relativistic three-particle equations allows us to calculate the
 mass spectrum of S-wave charmed baryons.

 The fact, that the forces of interaction for light quarks in charmed
 baryons $ g_1 $ and $ g_0 $ is actually such, as well as for diquarks in the
 bootstrap quark model for the usual baryons. Because of the
 rules of $1/N_c $-expansion the diquark forces are determined by
 one-gluon exchange.

 In the framework of suggested method it is possible to calculate the
 electrical formfactors and charge radii of charmed baryons,
 because we found the three-quark amplitudes $A_0 $ and $A_1 $.

\vspace{0.6cm}
\noindent
{\Large \bf Acknowledgements}
\vspace{0.6cm}

\noindent
 The authors would like to thank A.A.Andrianov, V.A.Franke, Yu.V.Novozhilov
 for useful discussions.

\newpage
\begin{center}
 Table 1.

 Coefficient of Chew-Mandelstam functions.
\vspace{0.3cm}

\large
\begin{tabular}{|c|c|c|c|} \hline
$J^P $  &$\gamma (J)$ &$\beta (J) $          & $ \delta (J)$    \\      \hline
$0^{+}$ &$\frac{1}{2}$ &$-\frac{1}{2}\frac{{(m_i - m_k)}^2}{{(m_i + m_k)}^2}$& 0  \\  \hline
$1^{+}$ &$\frac{1}{3}$ &$\frac{1}{6}\left (\frac{8 m_i m_k}{{(m_i +m_k)}^2}-1 \right )$ &$ -\frac{1}{6} {(m_i -m_k)}^2 $ \\
\hline
\end{tabular}
\end{center}
\normalsize

\vspace{1cm}
\begin{center}
 Table 2.

 Masses of charmed baryons of the multiplets
 $ J^{P}= {\frac{1}{2}}^{+}, {\frac{3}{2}}^{+} $.
\vspace{0.3cm}

\begin{tabular}{|l|l|l|l|l|}             \hline
 Quark       & $J^{P}={\frac{1}{2}}^{+}$ & M(GeV)       & $J^{P}={\frac{3}{2}}^{+}$     & M(GeV) \\
 contents    &                           &              &                         &        \\
 \hline
 udc         & $\Lambda_{c}^{+}$         & 2.284(2.285) &     ------              & ------
 \\ \hline
 uuc,udc,ddc & $\Sigma_{c}^{++,+,0}$     & 2.458(2.455) & $\Sigma_{c}^{* ++,+,0}$ & 2.516(2.519)
 \\ \hline
 usc,dsc     & $\Xi_{c}^{+,0(A)}$        & 2.467(2.467) & $\Xi_{c}^{* +,0}$       & 2.725(2.645)
 \\ \hline
 usc,dsc     & $\Xi_{c}^{+,0(S)}$        & 2.565(2.562) &     ------              & ------
 \\ \hline
 ssc         & $\Omega_{c}^{0}$          & 2.806(2.704) & $\Omega_{c}^{* 0}$      & 3.108
 \\ \hline
 ccu,ccd     & $\Omega_{ccq}^{++,+}$     & 3.527        & $\Omega_{q}^{* ++,+}$   & 3.597
 \\ \hline
 ccs         & $\Omega_{ccs}^{+}$        & 3.598        & $\Omega_{s}^{* +}$      & 3.700
 \\ \hline
 ccc         &      ------               & ------       & $\Omega_{c}^{++}$       & 4.792
 \\ \hline
\end{tabular}
\end{center}

\noindent
 The parameters of model: $ {\lambda}_q =10.7 $ and $ {\lambda}_c =6.5 $ are
 the cut-off parameter for light quarks and c-quark
 respectively, $g_0 =0.70 $ and $g_1 =0.56 $ are the vertex constant for
 light diquarks with $J^P = 0^{+}, 1^{+}$, $g_c =0.857$  for heavy diquarks.
 Quark masses $m=$0.495~GeV, $m_s =$0.77~GeV, $m_c =$1.655~GeV. The experimental
 values of the baryon masses are given in parentheses [33].

\newpage

\appendix

\section{Three-quark integral equations for the S-wave baryon amplitudes}

\vspace{0.5cm}

\centerline{\bf Multiplet $ J^P ={\frac{3}{2}}^{+}$.}
\vspace{0.5cm}

\noindent
1. ${\Sigma}_c^{* ++,+,0}$-baryons: $|{\Sigma}_c^{*} \rangle =|\{ q\uparrow
`q\uparrow c\uparrow \} \rangle $, where $q=u,d$.

$$A_1 (s,s_{12})=\hat{\alpha} b_1 (s_{12}) L_1 (s_{12})
+K_1 (s_{12}) [A_1^{(c)}(s,s_{13}^{\prime})+ A_1^{(c)}(s,s_{23}^{\prime})] $$
\hfill (A.1)
$$A_1^{(c)} (s,s_{13})=\hat{\alpha} b_1^{(c)} (s_{13}) L_1^{(c)} (s_{13})
+K_1^{(c)} (s_{13}) [A_1 (s,s_{12}^{\prime})+ A_1^{(c)}(s,s_{23}^{\prime})] $$
\vspace{0.1cm}

\noindent
2. ${\Xi}_c^{* +,0}$-baryons: $|{\Xi}_c^{*} \rangle =|\{ q\uparrow
s\uparrow c\uparrow \} \rangle $, where $q=u,d$.

$$A_1^{(s)} (s,s_{12})=\hat{\alpha} b_1^{(s)} (s_{12}) L_1^{(s)} (s_{12})
+K_1^{(s)} (s_{12}) [A_1^{(c)} (s,s_{13}^{\prime})+
 A_1^{(sc)}(s,s_{23}^{\prime})] $$
\hfill (A.2)
$$A_1^{(c)} (s,s_{13})=\hat{\alpha} b_1^{(c)} (s_{13}) L_1^{(c)} (s_{13})
+K_1^{(c)} (s_{13}) [A_1^{(s)} (s,s_{12}^{\prime})+
A_1^{(sc)}(s,s_{23}^{\prime})]  $$

$$A_1^{(sc)} (s,s_{23})=\hat{\alpha} b_1^{(sc)} (s_{23}) L_1^{(sc)} (s_{23})
+K_1^{(sc)} (s_{23}) [A_1^{(s)} (s,s_{12}^{\prime})+
A_1^{(c)}(s,s_{13}^{\prime})] $$
\vspace{0.1cm}

\noindent
3. ${\Omega}_c^{* 0}$-baryon $|{\Omega}_c^{* 0} \rangle =|\{ c\uparrow
s\uparrow s\uparrow \} \rangle $. The equations are analogous to (A.1) with
replacement $ A \to A^{(ss)}, A^{(c)} \to A^{(sc)} $.
\vspace{0.3cm}

\noindent
4. ${\Omega}_c^{* ++,+}$-baryons: $|{\Omega}_c^{* ++,+} \rangle =|\{ q\uparrow
c\uparrow c\uparrow \} \rangle ,q=u,d $. The equations are analogous to (A.1)
with replacement $ A \to A^{(cc)} $.
\vspace{0.3cm}

\noindent
5. ${\Omega}_s^{* +}$-baryon: $|{\Omega}_s^{* +} \rangle =|\{ s\uparrow
c\uparrow c\uparrow \} \rangle $. The equations are analogous to (A.1)
with replacement $ A \to A^{(cc)}, A^{(c)} \to A^{(sc)} $.
\vspace{0.3cm}

\noindent
6. ${\Omega}_c^{++}$-baryon: $|{\Omega}_c^{++} \rangle =|\{ c\uparrow
c\uparrow c\uparrow \} \rangle $.

$$ A_1^{(cc)} (s,s_{12})=\hat{\alpha} b_1^{(cc)}
(s_{12}) L_1^{(cc)} (s_{12})+K_1^{(cc)} (s_{12}) [A_1^{(cc)}
(s,s_{13}^{\prime})+ A_1^{(cc)}(s,s_{23}^{\prime})] \hspace{1.4cm}
{\rm (A.3)}  $$ \vspace{0.2cm}

\centerline{\bf Multiplet $ J^P ={\frac{1}{2}}^{+}$.}
\vspace{0.5cm}

\noindent
1. ${\Sigma}_c^{ ++,+,0}$-baryons: $|{\Sigma}_c^{+} \rangle =\sqrt{\frac{2}{3}}
|\{ u\uparrow d\uparrow c\downarrow \} \rangle -\frac{1}{\sqrt{6}}
|\{ u\uparrow d\downarrow c\uparrow \} \rangle -\frac{1}{\sqrt{6}}
|\{ u\downarrow d\uparrow c\uparrow \} \rangle $.
The spin-flavour parts of the wave functions for ${\Sigma}_c^{++}$ and ${\Sigma}_c^0 $-
baryons are derived with replacement $(ud) \to (uu) $ and $(ud) \to (dd) $
respectively.

$$ A_{1}(s,s_{12})=\hat{\alpha} b_{1}(s_{12}) L_{1}(s_{12})
+ K_{1}(s_{12}) [\frac{1}{4} A_{1}^{(c)}(s,s_{13}^{\prime})+
\frac{3}{4} A_{0}^{(c)}(s,s_{13}^{\prime})+ \frac{1}{4} A_{1}^{(c)}(s,s_{23}
^{\prime})+ \frac{3}{4} A_{0}^{(c)}(s,s_{23}^{\prime})]  $$

$$ A_{1}^{(c)}(s,s_{13})=\hat{\alpha} b_{1}^{(c)}(s_{13})
L_{1}^{(c)}(s_{13})+ K_{1}^{(c)}(s_{13})
[\frac{1}{2} A_{1}(s,s_{12}^{\prime})+
\frac{3}{4} A_{0}^{(c)}(s,s_{12}^{\prime})- \frac{1}{4} A_{1}^{(c)}(s,s_{12}
^{\prime})+ $$
$$+ \frac{1}{2} A_{1}(s,s_{23}^{\prime})+ \frac{3}{4} A_{0}^{(c)}
(s,s_{23}^{\prime})- \frac{1}{4} A_{1}^{(c)}(s,s_{23}^{\prime})]  $$
\hfill (A.4)
$$ A_{0}^{(c)}(s,s_{23})=\hat{\alpha} b_{0}^{(c)}(s_{23})
L_{0}^{(c)}(s_{23})+ K_{0}^{(c)}(s_{23})
 [\frac{1}{2} A_{1}(s,s_{13}^{\prime})+
\frac{1}{4} A_{1}^{(c)}(s,s_{13}^{\prime})+ \frac{1}{4} A_{0}^{(c)}(s,s_{13}
^{\prime})+ $$
$$+ \frac{1}{2} A_{1}(s,s_{12}^{\prime})+ \frac{1}{4} A_{1}^{(c)}
(s,s_{12}^{\prime})+ \frac{1}{4} A_{0}^{(c)}(s,s_{12}^{\prime})] $$
\vspace{0.1cm}

\noindent
2. ${\Lambda}_c^{+}$-baryon: $|{\Lambda}_c^{+} \rangle =\frac{1}{\sqrt{2}}
|\{ u\uparrow d\downarrow c\uparrow \} \rangle -\frac{1}{\sqrt{2}}
|\{ u\downarrow d\uparrow c\uparrow \} \rangle $

$$ A_{0}(s,s_{12})=\hat{\alpha} b_{0}(s_{12}) L_{0}(s_{12})
+ K_{0}(s_{12}) [\frac{1}{4} A_{0}^{(c)}(s,s_{13}^{\prime})+
\frac{3}{4} A_{1}^{(c)}(s,s_{13}^{\prime})+ \frac{1}{4} A_{0}^{(c)}(s,s_{23}
^{\prime})+ \frac{3}{4} A_{1}^{(c)}(s,s_{23}^{\prime})]  $$

$$ A_{1}^{(c)}(s,s_{13})=\hat{\alpha} b_{1}^{(c)}(s_{13})
L_{1}^{(c)}(s_{13})+ K_{1}^{(c)}(s_{13})
 [\frac{1}{2} A_{0}(s,s_{12}^{\prime})+
\frac{1}{4} A_{0}^{(c)}(s,s_{12}^{\prime})+ \frac{1}{4} A_{1}^{(c)}(s,s_{12}
^{\prime})+ $$
$$+ \frac{1}{2} A_{0}(s,s_{23}^{\prime})+ \frac{1}{4} A_{0}^{(c)}
(s,s_{23}^{\prime})+ \frac{1}{4} A_{1}^{(c)}(s,s_{23}^{\prime})]  $$
\hfill (A.5)
$$ A_{0}^{(c)}(s,s_{23})=\hat{\alpha} b_{0}^{(c)}(s_{23})
L_{0}^{(c)}(s_{23})+ K_{0}^{(c)}(s_{23})
 [\frac{1}{2} A_{0}(s,s_{12}^{\prime})+
\frac{3}{4} A_{1}^{(c)}(s,s_{12}^{\prime})- \frac{1}{4} A_{0}^{(c)}(s,s_{12}
^{\prime})+ $$
$$+ \frac{1}{2} A_{0}(s,s_{13}^{\prime})+ \frac{3}{4} A_{1}^{(c)}
(s,s_{13}^{\prime})- \frac{1}{4} A_{0}^{(c)}(s,s_{13}^{\prime})] $$
\vspace{0.1cm}

\noindent
3. ${\Xi}_c^{+,0 (A)}$-baryons: $|{\Xi}_c^{+ (A)} \rangle =\frac{1}{\sqrt{2}}
|\{ c\uparrow s\uparrow u\downarrow \} \rangle -\frac{1}{\sqrt{2}}
|\{ c\uparrow s\downarrow u\uparrow \} \rangle $
The spin-flavour part of the wave function for ${\Xi}_c^{0 (A)}$-baryon
are derived with replacement $(su) \to (sd) $.

$$ A_{0}^{(s)}(s,s_{12})=\hat{\alpha} b_{0}^{(s)}(s_{12})
L_{0}^{(s)}(s_{12})+ K_{0}^{(s)}(s_{12})
 [\frac{3}{8} A_{1}^{(c)}(s,s_{13}^{\prime})+
\frac{1}{8} A_{0}^{(c)}(s,s_{13}^{\prime})+ \frac{3}{8} A_{1}^{(sc)}(s,s_{13}
^{\prime})+  $$
$$+\frac{1}{8} A_{0}^{(sc)}(s,s_{13}^{\prime}) +\frac{3}{8} A_{1}^{(c)}
(s,s_{23}^{\prime})+ \frac{1}{8} A_{0}^{(c)}(s,s_{23}^{\prime})+ \frac{3}{8}
A_{1}^{(sc)}(s,s_{23}^{\prime})+ \frac{1}{8} A_{0}^{(sc)}(s,s_{23}^{\prime})] $$

$$ A_{1}^{(c)}(s,s_{13})=\hat{\alpha} b_{1}^{(c)}(s_{13})
L_{1}^{(c)}(s_{13})+ K_{1}^{(c)}(s_{13})
 [\frac{1}{2} A_{0}^{(s)}(s,s_{12}^{\prime})+
\frac{1}{4} A_{1}^{(sc)}(s,s_{12}^{\prime})+ \frac{1}{4} A_{0}^{(sc)}(s,s_{12}
^{\prime})+ $$
$$+ \frac{1}{2} A_{0}^{(s)}(s,s_{23}^{\prime})+ \frac{1}{4} A_{1}^{(sc)}
(s,s_{23}^{\prime})+ \frac{1}{4} A_{0}^{(sc)}(s,s_{23}^{\prime})]  $$

$$ A_{0}^{(c)}(s,s_{13})=\hat{\alpha} b_{0}^{(c)}(s_{13})
L_{0}^{(c)}(s_{13})+ K_{0}^{(c)}(s_{13})
 [\frac{1}{2} A_{0}^{(s)}(s,s_{12}^{\prime})+
\frac{3}{4} A_{1}^{(sc)}(s,s_{12}^{\prime})- \frac{1}{4} A_{0}^{(sc)}(s,s_{12}
^{\prime})+ $$
$$+ \frac{1}{2} A_{0}^{(s)}(s,s_{23}^{\prime})+ \frac{3}{4} A_{1}^{(sc)}
(s,s_{23}^{\prime})- \frac{1}{4} A_{0}^{(sc)}(s,s_{23}^{\prime})]  $$
\hfill (A.6)
$$ A_{1}^{(sc)}(s,s_{23})=\hat{\alpha} b_{1}^{(sc)}(s_{23})
L_{1}^{(sc)}(s_{23})+ K_{1}^{(sc)}(s_{23})
 [\frac{1}{2} A_{0}^{(s)}(s,s_{12}^{\prime})+
\frac{1}{4} A_{1}^{(c)}(s,s_{12}^{\prime})+ \frac{1}{4} A_{0}^{(c)}(s,s_{12}
^{\prime})+ $$
$$+ \frac{1}{2} A_{0}^{(s)}(s,s_{13}^{\prime})+ \frac{1}{4} A_{1}^{(c)}
(s,s_{13}^{\prime})+ \frac{1}{4} A_{0}^{(c)}(s,s_{13}^{\prime})]  $$

$$ A_{0}^{(sc)}(s,s_{23})=\hat{\alpha} b_{0}^{(sc)}(s_{23})
L_{0}^{(sc)}(s_{23})+ K_{0}^{(sc)}(s_{23})
 [\frac{1}{2} A_{0}^{(s)}(s,s_{12}^{\prime})+
\frac{3}{4} A_{1}^{(c)}(s,s_{12}^{\prime})- \frac{1}{4} A_{0}^{(c)}(s,s_{12}
^{\prime})+ $$
$$+ \frac{1}{2} A_{0}^{(s)}(s,s_{13}^{\prime})+ \frac{3}{4} A_{1}^{(c)}
(s,s_{13}^{\prime})- \frac{1}{4} A_{0}^{(c)}(s,s_{13}^{\prime})]  $$
\vspace{0.1cm}

\noindent
4. ${\Xi}_c^{+,0 (S)}$-baryons:$|{\Xi}_c^{+ (S)} \rangle =\sqrt{\frac{2}{3}}
|\{ u\uparrow s\uparrow c\downarrow \}\rangle -\frac{1}{\sqrt{6}}
|\{ u\uparrow s\downarrow c\uparrow \} \rangle -\frac{1}{\sqrt{6}}
|\{ u\downarrow s\uparrow c\uparrow \}\rangle $.
The spin-flavour part of the wave function for ${\Xi}_c^{0 (S)}$-baryon
are derived with replacement $(us) \to (ds) $.

$$ A_{1}^{(s)}(s,s_{12})=\hat{\alpha} b_{1}^{(s)}(s_{12})
L_{1}^{(s)}(s_{12})+ K_{1}^{(s)}(s_{12})
 [\frac{1}{8} A_{1}^{(c)}(s,s_{13}^{\prime})+
\frac{3}{8} A_{0}^{(c)}(s,s_{13}^{\prime})+ \frac{1}{8} A_{1}^{(sc)}(s,s_{13}
^{\prime})+  $$
$$+\frac{3}{8} A_{0}^{(sc)}(s,s_{13}^{\prime}) +\frac{1}{8} A_{1}^{(c)}
(s,s_{23}^{\prime})+ \frac{3}{8} A_{0}^{(c)}(s,s_{23}^{\prime})+ \frac{1}{8}
A_{1}^{(sc)}(s,s_{23}^{\prime})+ \frac{3}{8} A_{0}^{(sc)}(s,s_{23}^{\prime})] $$

$$ A_{1}^{(c)}(s,s_{13})=\hat{\alpha} b_{1}^{(c)}(s_{13})
L_{1}^{(c)}(s_{13})+ K_{1}^{(c)}(s_{13})
 [\frac{1}{2} A_{1}^{(s)}(s,s_{12}^{\prime})-
\frac{1}{4} A_{1}^{(sc)}(s,s_{12}^{\prime})+ \frac{3}{4} A_{0}^{(sc)}(s,s_{12}
^{\prime})+ $$
$$+ \frac{1}{2} A_{1}^{(s)}(s,s_{23}^{\prime})- \frac{1}{4} A_{1}^{(sc)}
(s,s_{23}^{\prime})+ \frac{3}{4} A_{0}^{(sc)}(s,s_{23}^{\prime})]  $$

$$ A_{0}^{(c)}(s,s_{13})=\hat{\alpha} b_{0}^{(c)}(s_{13})
L_{0}^{(c)}(s_{13})+ K_{0}^{(c)}(s_{13})
 [\frac{1}{2} A_{1}^{(s)}(s,s_{12}^{\prime})+
\frac{1}{4} A_{1}^{(sc)}(s,s_{12}^{\prime})+ \frac{1}{4} A_{0}^{(sc)}(s,s_{12}
^{\prime})+ $$
$$+ \frac{1}{2} A_{1}^{(s)}(s,s_{23}^{\prime})+ \frac{1}{4} A_{1}^{(sc)}
(s,s_{23}^{\prime})+ \frac{1}{4} A_{0}^{(sc)}(s,s_{23}^{\prime})]  $$
\hfill (A.7)
$$ A_{1}^{(sc)}(s,s_{23})=\hat{\alpha} b_{1}^{(sc)}(s_{23})
L_{1}^{(sc)}(s_{23})+ K_{1}^{(sc)}(s_{23})
 [\frac{1}{2} A_{1}^{(s)}(s,s_{12}^{\prime})-
\frac{1}{4} A_{1}^{(c)}(s,s_{12}^{\prime})+ \frac{3}{4} A_{0}^{(c)}(s,s_{12}
^{\prime})+ $$
$$+ \frac{1}{2} A_{1}^{(s)}(s,s_{13}^{\prime})- \frac{1}{4} A_{1}^{(c)}
(s,s_{13}^{\prime})+ \frac{3}{4} A_{0}^{(c)}(s,s_{13}^{\prime})]  $$

$$ A_{0}^{(sc)}(s,s_{23})=\hat{\alpha} b_{0}^{(sc)}(s_{23})
L_{0}^{(sc)}(s_{23})+ K_{0}^{(sc)}(s_{23})
 [\frac{1}{2} A_{1}^{(s)}(s,s_{12}^{\prime})+
\frac{1}{4} A_{1}^{(c)}(s,s_{12}^{\prime})+ \frac{1}{4} A_{0}^{(c)}(s,s_{12}
^{\prime})+ $$
$$+ \frac{1}{2} A_{1}^{(s)}(s,s_{13}^{\prime})+ \frac{1}{4} A_{1}^{(c)}
(s,s_{13}^{\prime})+ \frac{1}{4} A_{0}^{(c)}(s,s_{13}^{\prime})]  $$
\vspace{0.1cm}

\noindent
5. ${\Omega}_c^0 $-baryon: $|{\Omega}_c^0 \rangle =\sqrt{\frac{2}{3}}
|\{ s\uparrow s\uparrow c\downarrow \}\rangle -\frac{1}{\sqrt{3}}
|\{ s\uparrow s\downarrow c\uparrow \} \rangle  $.
The equations are analogous to (A.4) with replacement $ A\to A^{(ss)}, A^{(c)} \to A^{(sc)}$.
\vspace{0.5cm}

\noindent
6. ${\Omega}_{ccq}^{++,+}$-baryons: $|{\Omega}_{ccq}^{++} \rangle =\sqrt{\frac{2}{3}}
|\{ c\uparrow c\uparrow u\downarrow \}\rangle -\frac{1}{\sqrt{3}}
|\{ c\uparrow c\downarrow u\uparrow \} \rangle  $.
The spin-flavour part of the wave function for ${\Omega}_{ccq}^{+}$-particle
are derived with replacement $u\to d$. The system of integral equations is
analogous to (A.4) with replacement $ A\to A^{(cc)}$.
\vspace{0.5cm}

\noindent
7. ${\Omega}_{ccs}^{+}$-baryon: $|{\Omega}_{ccs}^{+} \rangle =\sqrt{\frac{2}{3}}
|\{ c\uparrow c\uparrow s\downarrow \}\rangle -\frac{1}{\sqrt{3}}
|\{ c\uparrow c\downarrow s\uparrow \} \rangle  $.
The system of integral equations is analogous to (A.4)
with replacement $ A\to A^{(cc)}, A^{(c)} \to A^{(sc)} $.
\vspace{0.5cm}



\section{The systems of approximate algebraic equations for multiplets of
 charmed baryons $ J^P ={\frac{1}{2}}^{+}, J^P ={\frac{3}{2}}^{+} $}
\vspace{0.3cm}

\centerline{\bf Multiplet $ J^P ={\frac{3}{2}}^{+}$.}
\vspace{0.5cm}

\noindent
1. ${\Sigma}_c^{* ++,+,0}$-baryons:

$$ {\alpha}_1 (s,s_0)=\hat{\alpha}+ 2{\alpha}_1^{(c)}(s,s_0) I_{1,1}(s,s_0)
 \frac{b_1^{(c)}(s_0)}{b_1 (s_0)}   $$
\hfill (B.1)
$$ {\alpha}_1^{(c)}(s,s_0)=\hat{\alpha}+ {\alpha}_1 (s,s_0)
 I_{1,1}^{(c)}(s,s_0) \frac{b_1 (s_0)}{b_1^{(c)} (s_0)} +
 {\alpha}_1^{(c)}(s,s_0) I_{1,1}^{(c)}(s,s_0)  $$
\vspace{0.5cm}

\noindent
2. ${\Xi}_c^{* +,0}$-baryons:

$$ {\alpha}_1^{(s)}(s,s_0)=\hat{\alpha}+{\alpha}_1^{(c)}(s,s_0) I_{1,1}^{(s)}
(s,s_0)\frac{b_1^{(c)}(s_0)}{b_1^{(s)}(s_0)}+
{\alpha}_1^{(sc)}(s,s_0) I_{1,1}^{(s)}(s,s_0)  $$
\hfill (B.2)
$$ {\alpha}_1^{(c)}(s,s_0)=\hat{\alpha}+{\alpha}_1^{(s)}(s,s_0) I_{1,1}^{(c)}
(s,s_0)\frac{b_1^{(s)}(s_0)}{b_1^{(c)}(s_0)}+
{\alpha}_1^{(sc)}(s,s_0) I_{1,1}^{(c)}(s,s_0)
\frac{b_1^{(sc)}(s_0)}{b_1^{(c)}(s_0)}    $$

$$ {\alpha}_1^{(sc)}(s,s_0)=\hat{\alpha}+{\alpha}_1^{(s)}(s,s_0) I_{1,1}^{(sc)}
(s,s_0)\frac{b_1^{(s)}(s_0)}{b_1^{(sc)}(s_0)}+
{\alpha}_1^{(c)}(s,s_0) I_{1,1}^{(sc)}(s,s_0)
\frac{b_1^{(c)}(s_0)}{b_1^{(sc)}(s_0)}  $$
\vspace{0.5cm}

\noindent
3. ${\Omega}_c^{* 0}$-baryon: The equations are analogous to (A.8) with replacement
$ {\alpha} \to {\alpha}^{(ss)}, {\alpha}^{(c)} \to {\alpha}^{(sc)},
b^{(c)} \to b^{(sc)}, b \to b^{(ss)} $.
\vspace{0.5cm}

\noindent
4. ${\Omega}_c^{* ++,+}$-baryons: The equations are analogous to (A.8) with
replacement $ {\alpha} \to {\alpha}^{(cc)}, b \to b^{(cc)} $.
\vspace{0.5cm}

\noindent
5. ${\Omega}_s^{* +}$-baryon: The equations are analogous to (A.8) with replacement
$ {\alpha} \to {\alpha}^{(cc)}, {\alpha}^{(c)} \to {\alpha}^{(sc)},
b \to b^{(cc)}, b^{(c)} \to b^{(sc)} $.
\vspace{0.5cm}

\noindent
6. ${\Omega}_c^{++}$-baryon:

\vspace{0.3cm}
$$ {\alpha}_1^{(cc)}(s,s_0)=
\hat{\alpha} + {\alpha}_1^{(cc)}(s,s_0) 2 I_{1,1}^{(cc)}(s,s_0)
\hspace{2.5cm}  {\rm (B.3)}  $$
\vspace{1.8cm}

\centerline{\bf Multiplet $ J^P ={\frac{1}{2}}^{+}$.}
\vspace{0.5cm}

\noindent
1. ${\Sigma}_c^{ ++,+,0}$-baryons:

$${\alpha}_1 (s,s_0)=\hat{\alpha}+\frac{1}{2} {\alpha}_1^{(c)}(s,s_0)
I_{1,1}(s,s_0)\frac{b_1^{(c)}(s_0)}{b_1 (s_0)}+
\frac{3}{2} {\alpha}_0^{(c)}(s,s_0)
I_{1,0}(s,s_0)\frac{b_0^{(c)}(s_0)}{b_1 (s_0)}  $$

$${\alpha}_1^{(c)} (s,s_0)=\hat{\alpha}+{\alpha}_1 (s,s_0) I_{1,1}^{(c)}
(s,s_0)\frac{b_1 (s_0)}{b_1^{(c)}(s_0)}+  $$
$$+\frac{3}{2} {\alpha}_0^{(c)}(s,s_0)
I_{1,0}^{(c)}(s,s_0)\frac{b_0^{(c)}(s_0)}{b_1^{(c)} (s_0)}
-\frac{1}{2} {\alpha}_1^{(c)}(s,s_0)I_{1,1}^{(c)}(s,s_0)  $$
\hfill (B.4)
$${\alpha}_0^{(c)} (s,s_0)=\hat{\alpha}+{\alpha}_1 (s,s_0) I_{0,1}^{(c)}
(s,s_0)\frac{b_1 (s_0)}{b_0^{(c)}(s_0)}+  $$
$$+\frac{1}{2} {\alpha}_1^{(c)}(s,s_0)I_{0,1}^{(c)}(s,s_0)
\frac{b_1^{(c)}(s_0)}{b_0^{(c)}(s_0)} +
\frac{1}{2} {\alpha}_0^{(c)}(s,s_0) I_{0,0}^{(c)}(s,s_0) $$
\vspace{0.5cm}

\noindent
2. ${\Lambda}_c^{+}$-baryon:

$${\alpha}_0 (s,s_0)=\hat{\alpha}+\frac{1}{2} {\alpha}_0^{(c)}(s,s_0)
I_{0,0}(s,s_0)\frac{b_0^{(c)}(s_0)}{b_0 (s_0)}+
\frac{3}{2} {\alpha}_1^{(c)}(s,s_0)
I_{0,1}(s,s_0)\frac{b_1^{(c)}(s_0)}{b_0 (s_0)}  $$

$${\alpha}_1^{(c)} (s,s_0)=\hat{\alpha}+{\alpha}_0 (s,s_0) I_{1,0}^{(c)}
(s,s_0)\frac{b_0 (s_0)}{b_1^{(c)}(s_0)}+  $$
$$+\frac{1}{2} {\alpha}_1^{(c)}(s,s_0)I_{1,1}^{(c)}(s,s_0)+
\frac{1}{2} {\alpha}_0^{(c)}(s,s_0)
I_{1,0}^{(c)}(s,s_0)\frac{b_0^{(c)}(s_0)}{b_1^{(c)} (s_0)}  $$
\hfill (B.5)
$${\alpha}_0^{(c)} (s,s_0)=\hat{\alpha}+{\alpha}_0 (s,s_0) I_{0,0}^{(c)}
(s,s_0)\frac{b_0 (s_0)}{b_0^{(c)}(s_0)}+  $$
$$+\frac{3}{2} {\alpha}_1^{(c)}(s,s_0)I_{0,1}^{(c)}(s,s_0)
\frac{b_1^{(c)}(s_0)}{b_0^{(c)}(s_0)} -
\frac{1}{2} {\alpha}_0^{(c)}(s,s_0) I_{0,0}^{(c)}(s,s_0) $$
\vspace{0.5cm}

\noindent
3. ${\Xi}_c^{+,0 (A)}$-baryons:

$${\alpha}_0^{(s)}(s,s_0)=\hat{\alpha}+\frac{3}{4}{\alpha}_1^{(c)}(s,s_0)
I_{0,1}^{(s)}(s,s_0)\frac{b_1^{(c)}(s_0)}{b_0^{(s)}(s_0)}+
\frac{1}{4}{\alpha}_0^{(c)}(s,s_0)
I_{0,0}^{(s)}(s,s_0)\frac{b_0^{(c)}(s_0)}{b_0^{(s)}(s_0)}+  $$
$$+\frac{3}{4}{\alpha}_1^{(sc)}(s,s_0)
I_{0,1}^{(s)}(s,s_0)\frac{b_1^{(sc)}(s_0)}{b_0^{(s)}(s_0)}+
\frac{1}{4}{\alpha}_0^{(sc)}(s,s_0)
I_{0,0}^{(s)}(s,s_0)\frac{b_0^{(sc)}(s_0)}{b_0^{(s)}(s_0)} $$

$${\alpha}_1^{(c)}(s,s_0)=\hat{\alpha}+{\alpha}_0^{(s)}(s,s_0) I_{1,0}^{(c)}
(s,s_0) \frac{b_0^{(s)}(s_0)}{b_1^{(c)}(s_0)}+  $$
$$+\frac{1}{2} {\alpha}_1^{(sc)}(s,s_0) I_{1,1}^{(c)}(s,s_0)
\frac{b_1^{(sc)}(s_0)}{b_1^{(c)}(s_0)}+
\frac{1}{2} {\alpha}_0^{(sc)}(s,s_0) I_{1,0}^{(c)}(s,s_0)
\frac{b_0^{(sc)}(s_0)}{b_1^{(c)}(s_0)}  $$

$${\alpha}_0^{(c)}(s,s_0)=\hat{\alpha}+{\alpha}_0^{(s)}(s,s_0) I_{0,0}^{(c)}
(s,s_0)\frac{b_0^{(s)}(s_0)}{b_0^{(c)}(s_0)}+  $$
$$+\frac{3}{2} {\alpha}_1^{(sc)}(s,s_0) I_{0,1}^{(c)}(s,s_0)
\frac{b_1^{(sc)}(s_0)}{b_0^{(c)}(s_0)}-
\frac{1}{2} {\alpha}_0^{(sc)}(s,s_0) I_{0,0}^{(c)}(s,s_0)
\frac{b_0^{(sc)}(s_0)}{b_0^{(c)}(s_0)}  $$
\hfill (B.6)
$${\alpha}_1^{(sc)}(s,s_0)=\hat{\alpha}+{\alpha}_0^{(s)}(s,s_0) I_{1,0}^{(sc)}
(s,s_0)\frac{b_0^{(s)}(s_0)}{b_1^{(sc)}(s_0)}+  $$
$$+\frac{1}{2} {\alpha}_1^{(c)}(s,s_0) I_{1,1}^{(sc)}(s,s_0)
\frac{b_1^{(c)}(s_0)}{b_1^{(sc)}(s_0)}+
\frac{1}{2} {\alpha}_0^{(c)}(s,s_0) I_{1,0}^{(sc)}(s,s_0)
\frac{b_0^{(c)}(s_0)}{b_1^{(sc)}(s_0)}  $$

$${\alpha}_0^{(sc)}(s,s_0)=\hat{\alpha}+{\alpha}_0^{(s)}(s,s_0) I_{0,0}^{(sc)}
(s,s_0) \frac{b_0^{(s)}(s_0)}{b_0^{(sc)}(s_0)}+  $$
$$+\frac{3}{2} {\alpha}_1^{(c)}(s,s_0) I_{0,1}^{(sc)}(s,s_0)
\frac{b_1^{(c)}(s_0)}{b_0^{(sc)}(s_0)}-
\frac{1}{2} {\alpha}_0^{(c)}(s,s_0) I_{0,0}^{(sc)}(s,s_0)
\frac{b_0^{(c)}(s_0)}{b_0^{(sc)}(s_0)}  $$
\vspace{0.5cm}

\noindent
4. ${\Xi}_c^{+,0 (S)}$-baryons:

$${\alpha}_1^{(s)}(s,s_0)=\hat{\alpha}+\frac{1}{4}{\alpha}_1^{(c)}(s,s_0)
I_{1,1}^{(s)}(s,s_0)\frac{b_1^{(c)}(s_0)}{b_1^{(s)}(s_0)}+
\frac{3}{4}{\alpha}_0^{(c)}(s,s_0)
I_{1,0}^{(s)}(s,s_0)\frac{b_0^{(c)}(s_0)}{b_1^{(s)}(s_0)}+  $$
$$+\frac{1}{4}{\alpha}_1^{(sc)}(s,s_0)
I_{1,1}^{(s)}(s,s_0)\frac{b_1^{(sc)}(s_0)}{b_1^{(s)}(s_0)}+
\frac{3}{4}{\alpha}_0^{(sc)}(s,s_0)
I_{1,0}^{(s)}(s,s_0)\frac{b_0^{(sc)}(s_0)}{b_1^{(s)}(s_0)} $$

$${\alpha}_1^{(c)}(s,s_0)=\hat{\alpha}+{\alpha}_1^{(s)}(s,s_0) I_{1,1}^{(c)}
(s,s_0)\frac{b_1^{(s)}(s_0)}{b_1^{(c)}(s_0)}-  $$
$$-\frac{1}{2} {\alpha}_1^{(sc)}(s,s_0) I_{1,1}^{(c)}(s,s_0)
\frac{b_1^{(sc)}(s_0)}{b_1^{(c)}(s_0)}+
\frac{3}{2} {\alpha}_0^{(sc)}(s,s_0) I_{1,0}^{(c)}(s,s_0)
\frac{b_0^{(sc)}(s_0)}{b_1^{(c)}(s_0)}  $$

$${\alpha}_0^{(c)}(s,s_0)=\hat{\alpha}+{\alpha}_1^{(s)}(s,s_0) I_{0,1}^{(c)}
(s,s_0) \frac{b_1^{(s)}(s_0)}{b_0^{(c)}(s_0)}+  $$
$$+\frac{1}{2} {\alpha}_1^{(sc)}(s,s_0) I_{0,1}^{(c)}(s,s_0)
\frac{b_1^{(sc)}(s_0)}{b_0^{(c)}(s_0)}+
\frac{1}{2} {\alpha}_0^{(sc)}(s,s_0) I_{0,0}^{(c)}(s,s_0)
\frac{b_0^{(sc)}(s_0)}{b_0^{(c)}(s_0)}  $$
\hfill (B.7)
$${\alpha}_1^{(sc)}(s,s_0)=\hat{\alpha}+{\alpha}_1^{(s)}(s,s_0) I_{1,1}^{(sc)}
(s,s_0)\frac{b_1^{(s)}(s_0)}{b_1^{(sc)}(s_0)}-  $$
$$-\frac{1}{2} {\alpha}_1^{(c)}(s,s_0) I_{1,1}^{(sc)}(s,s_0)
\frac{b_1^{(c)}(s_0)}{b_1^{(sc)}(s_0)}+
\frac{3}{2} {\alpha}_0^{(c)}(s,s_0) I_{1,0}^{(sc)}(s,s_0)
\frac{b_0^{(c)}(s_0)}{b_1^{(sc)}(s_0)}  $$

$${\alpha}_0^{(sc)}(s,s_0)=\hat{\alpha}+{\alpha}_1^{(s)}(s,s_0) I_{0,1}^{(sc)}
(s,s_0)\frac{b_1^{(s)}(s_0)}{b_0^{(sc)}(s_0)}+  $$
$$+\frac{1}{2} {\alpha}_1^{(c)}(s,s_0) I_{0,1}^{(sc)}(s,s_0)
\frac{b_1^{(c)}(s_0)}{b_0^{(sc)}(s_0)}+
\frac{1}{2} {\alpha}_0^{(c)}(s,s_0) I_{0,0}^{(sc)}(s,s_0)
\frac{b_0^{(c)}(s_0)}{b_0^{(sc)}(s_0)}  $$
\vspace{0.5cm}

\noindent
5. ${\Omega}_c^{0}$-baryon: The equations are analogous to (B.4) with replacement
$ {\alpha} \to {\alpha}^{(ss)}, {\alpha}^{(c)} \to {\alpha}^{(sc)},
b^{(c)} \to b^{(sc)}, b \to b^{(ss)} $.
\vspace{0.5cm}

\noindent
6. ${\Omega}_{ccq}^{++,+}$-baryons: The equations are analogous to (B.4) with
replacement $ {\alpha} \to {\alpha}^{(cc)}, b \to b^{(cc)} $.
\vspace{0.5cm}

\noindent
7. ${\Omega}_{ccs}^{+}$-baryon: The equations are analogous to (B.4) with replacement
$ {\alpha} \to {\alpha}^{(cc)}, {\alpha}^{(c)} \to {\alpha}^{(sc)},
b \to b^{(cc)}, b^{(c)} \to b^{(sc)} $.
\vspace{0.5cm}

\newpage

\vspace*{0.3cm}
\noindent
\unitlength 1mm
\hspace{-0.4cm}
\begin{picture}(151.67,111.67)
\thicklines
\put(11.33,42.33){\circle{5.45}}
\put(15.00,46.67){\circle{5.45}}
\put(18.67,51.00){\circle{5.45}}
\put(20.67,52.67){\vector(1,1){6.00}}
\put(12.33,40.00){\vector(1,0){27.00}}
\put(21.00,52.33){\line(3,-5){7.40}}
\put(27.67,40.00){\vector(1,-1){8.67}}
\put(44.67,40.00){\makebox(0,0)[cc]{+}}
\put(51.33,40.00){\makebox(0,0)[cc]{...}}
\put(65.00,42.67){\circle{5.45}}
\put(68.67,47.00){\circle{5.45}}
\put(72.33,51.33){\circle{5.45}}
\put(74.33,53.00){\vector(1,1){6.00}}
\put(74.67,52.67){\line(3,-5){7.40}}
\put(66.00,40.33){\line(1,0){15.67}}
\put(83.67,38.33){\circle{5.45}}
\put(87.33,34.00){\circle{5.45}}
\put(91.00,29.67){\circle{5.45}}
\put(92.00,27.00){\vector(1,-4){2.17}}
\put(92.67,27.33){\vector(3,-2){7.00}}
\put(95.67,40.00){\makebox(0,0)[cc]{+}}
\put(3.33,40.00){\vector(1,0){6.00}}
\put(3.33,40.40){\vector(1,0){6.00}}
\put(57.00,40.00){\vector(1,0){7.00}}
\put(102.00,40.00){\vector(1,0){6.00}}
\put(57.00,40.40){\vector(1,0){7.00}}
\put(102.00,40.40){\vector(1,0){6.00}}
\put(109.00,42.67){\circle{5.45}}
\put(112.67,47.00){\circle{5.45}}
\put(116.33,51.33){\circle{5.45}}
\put(109.33,40.00){\line(2,-1){13.00}}
\put(122.00,33.67){\line(-1,6){3.00}}
\put(124.33,31.33){\circle{5.45}}
\put(128.00,27.00){\circle{5.45}}
\put(131.67,22.67){\circle{5.45}}
\put(119.33,51.67){\line(4,-1){15.00}}
\put(134.33,47.92){\line(0,-1){26.25}}
\put(134.67,50.67){\circle{5.45}}
\put(138.33,55.00){\circle{5.45}}
\put(142.00,59.33){\circle{5.45}}
\put(143.67,61.33){\vector(1,4){2.08}}
\put(144.00,61.67){\vector(2,1){7.67}}
\put(133.33,20.33){\vector(3,-4){4.75}}
\put(9.67,90.00){\vector(1,0){9.33}}
\put(9.67,90.40){\vector(1,0){9.33}}
\put(19.00,90.00){\vector(3,2){9.00}}
\put(19.33,90.00){\vector(1,0){11.67}}
\put(19.33,90.00){\vector(3,-2){8.67}}
\put(37.00,90.00){\makebox(0,0)[cc]{+}}
\put(43.33,90.00){\vector(1,0){9.33}}
\put(43.33,90.40){\vector(1,0){9.33}}
\put(55.33,92.67){\circle{7.36}}
\put(57.67,95.33){\vector(2,3){4.22}}
\put(58.33,95.00){\vector(4,1){7.67}}
\put(56.00,89.00){\vector(2,-1){9.67}}
\put(72.00,90.00){\makebox(0,0)[cc]{+}}
\put(77.67,90.00){\vector(1,0){9.33}}
\put(77.67,90.40){\vector(1,0){9.33}}
\put(87.67,92.67){\circle{5.45}}
\put(91.33,97.00){\circle{5.45}}
\put(95.00,101.33){\circle{5.45}}
\put(96.67,103.33){\vector(1,4){2.08}}
\put(97.00,103.67){\vector(2,1){7.67}}
\put(87.33,90.00){\vector(2,-1){12.00}}
\put(112.00,90.00){\makebox(0,0)[cc]{+}}
\put(18.67,77.00){\makebox(0,0)[cb]{a}}
\put(54.00,77.00){\makebox(0,0)[cb]{b}}
\put(88.00,77.00){\makebox(0,0)[cb]{c}}
\put(18.33,21.00){\makebox(0,0)[cb]{d}}
\put(75.00,21.00){\makebox(0,0)[cb]{e}}
\put(119.33,21.00){\makebox(0,0)[cb]{f}}
\end{picture}

\vspace*{-0.5cm}

\noindent
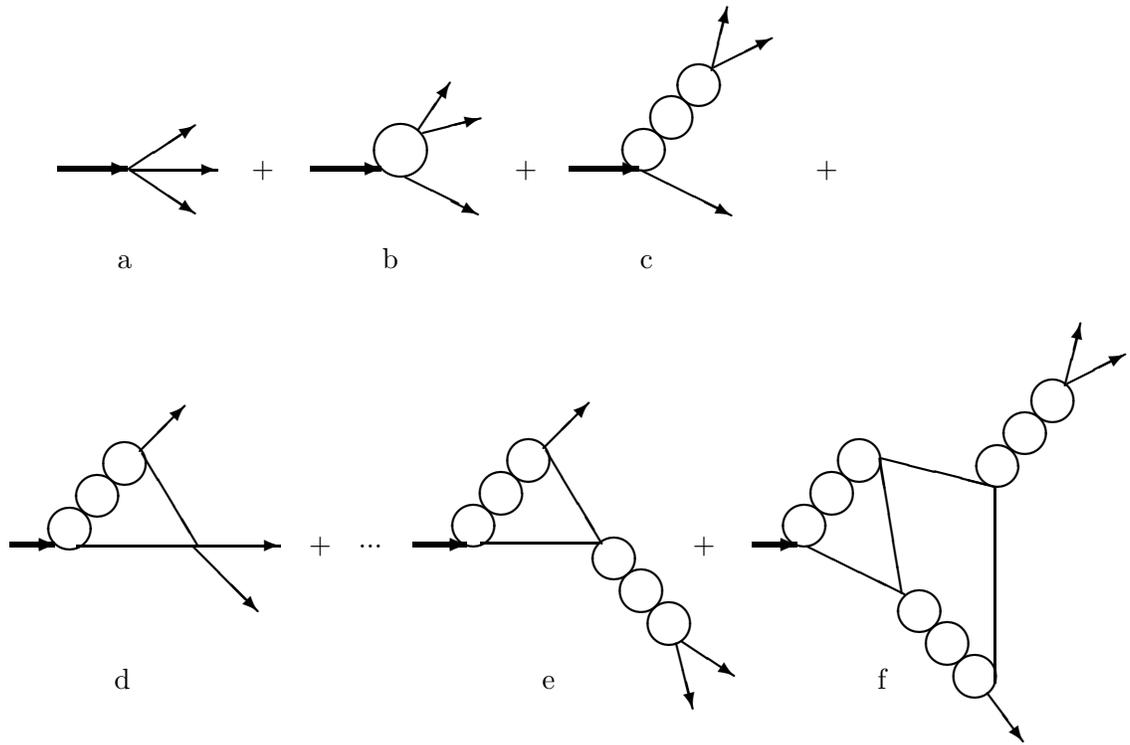
Fig.1. Diagrams which correspond to: a -- production of three quarks,
b--e -- subsequent pair interaction.

\vspace*{3.5cm}
\noindent
\unitlength 1mm
\hspace{-0.4cm}
\begin{picture}(142.67,47.00)
\thicklines
\put(5.00,30.00){\vector(1,0){9.67}}
\put(5.00,29.60){\vector(1,0){9.67}}
\put(14.67,29.67){\vector(2,1){15.33}}
\put(14.67,29.67){\vector(1,0){17.00}}
\put(15.00,29.67){\vector(3,-2){14.67}}
\put(40.33,30.00){\vector(1,0){10.27}}
\put(40.33,29.60){\vector(1,0){10.27}}
\put(53.33,29.00){\circle*{5.20}}
\put(52.67,26.67){\vector(2,-1){12.33}}
\put(56.33,32.67){\circle{4.71}}
\put(57.33,34.67){\vector(4,1){12.33}}
\put(70.33,29.33){\makebox(0,0)[cc]{+}}
\put(33.00,39.00){\makebox(0,0)[lb]{1}}
\put(32.67,29.33){\makebox(0,0)[lc]{2}}
\put(32.33,19.67){\makebox(0,0)[lc]{3}}
\put(37.33,29.67){\makebox(0,0)[cc]{+}}
\put(57.33,34.67){\vector(3,4){7.75}}
\put(75.33,30.00){\vector(1,0){10.27}}
\put(75.33,29.60){\vector(1,0){10.27}}
\put(110.33,30.00){\vector(1,0){10.27}}
\put(110.33,29.60){\vector(1,0){10.27}}
\put(88.33,29.00){\circle*{5.20}}
\put(123.33,29.00){\circle*{5.20}}
\put(87.67,26.67){\vector(2,-1){12.33}}
\put(122.67,26.67){\vector(2,-1){12.33}}
\put(91.33,32.67){\circle{4.71}}
\put(126.33,32.67){\circle{4.71}}
\put(92.33,34.67){\vector(4,1){12.33}}
\put(127.33,34.67){\vector(4,1){12.33}}
\put(105.33,29.67){\makebox(0,0)[cc]{+}}
\put(92.33,34.67){\vector(3,4){7.75}}
\put(127.33,34.67){\vector(3,4){7.75}}
\put(67.00,46.33){\makebox(0,0)[lb]{1}}
\put(72.33,38.00){\makebox(0,0)[lc]{2}}
\put(68.33,19.67){\makebox(0,0)[lc]{3}}
\put(102.33,47.00){\makebox(0,0)[lb]{1}}
\put(107.67,38.33){\makebox(0,0)[lc]{3}}
\put(103.00,20.00){\makebox(0,0)[lc]{2}}
\put(137.33,47.00){\makebox(0,0)[lb]{2}}
\put(142.67,38.67){\makebox(0,0)[lc]{3}}
\put(138.00,20.33){\makebox(0,0)[lc]{1}}
\end{picture}

\vspace*{-0.5cm}

\noindent
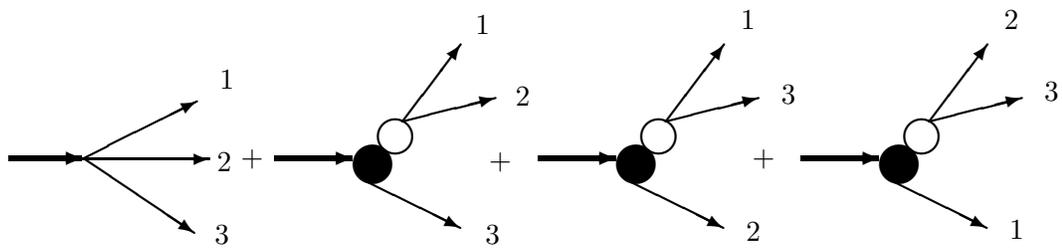
Fig.2. Combination of diagrams in accordance with the last
interaction of the particles.

\newpage

\vspace*{0.4cm}
\unitlength 1mm
\hspace{-1.0cm}
\begin{picture}(151.67,66.00)
\thicklines
\put(5.00,49.73){\vector(1,0){11.33}}
\put(5.00,49.33){\vector(1,0){11.33}}
\put(19.00,49.33){\circle*{5.20}}
\put(22.33,53.33){\circle{4.85}}
\put(18.67,46.67){\vector(4,-1){10.00}}
\put(23.00,55.67){\vector(3,4){6.50}}
\put(23.67,56.00){\vector(3,1){11.33}}
\put(31.33,66.00){\makebox(0,0)[lb]{1}}
\put(37.33,59.00){\makebox(0,0)[lc]{2}}
\put(31.00,44.00){\makebox(0,0)[lc]{3}}
\put(35.00,50.00){\makebox(0,0)[cc]{=}}
\put(39.00,50.00){\vector(1,0){11.00}}
\put(39.00,49.60){\vector(1,0){11.00}}
\put(53.00,50.00){\circle{6.00}}
\put(52.67,47.00){\vector(3,-1){10.67}}
\put(53.67,52.67){\vector(2,3){6.44}}
\put(54.33,53.00){\vector(3,1){12.33}}
\put(62.33,63.33){\makebox(0,0)[lb]{1}}
\put(69.33,57.33){\makebox(0,0)[lc]{2}}
\put(66.33,43.33){\makebox(0,0)[lc]{3}}
\put(71.33,50.00){\makebox(0,0)[cc]{+}}
\put(77.33,50.00){\vector(1,0){11.33}}
\put(77.33,49.60){\vector(1,0){11.33}}
\put(91.33,49.67){\circle*{5.20}}
\put(94.67,53.67){\circle{4.85}}
\put(91.67,47.00){\line(4,-3){8.33}}
\put(97.00,54.00){\line(1,-4){3.32}}
\put(97.33,53.33){\vector(2,3){6.89}}
\put(102.00,38.67){\circle{5.21}}
\put(104.33,37.00){\vector(2,-1){6.67}}
\put(104.33,36.67){\vector(1,-4){1.83}}
\put(105.00,66.00){\makebox(0,0)[cb]{3}}
\put(100.00,48.33){\makebox(0,0)[lc]{2}}
\put(113.33,33.33){\makebox(0,0)[lc]{2}}
\put(107.33,27.33){\makebox(0,0)[lt]{1}}
\put(107.67,49.67){\makebox(0,0)[cc]{+}}
\put(115.00,49.73){\vector(1,0){11.33}}
\put(115.00,49.33){\vector(1,0){11.33}}
\put(129.00,49.33){\circle*{5.20}}
\put(132.33,53.33){\circle{4.85}}
\put(129.33,46.67){\line(4,-3){8.33}}
\put(134.67,53.67){\line(1,-4){3.32}}
\put(135.00,53.00){\vector(2,3){6.89}}
\put(139.67,38.33){\circle{5.21}}
\put(142.00,36.67){\vector(2,-1){6.67}}
\put(142.00,36.33){\vector(1,-4){1.83}}
\put(142.33,66.00){\makebox(0,0)[cb]{3}}
\put(137.67,48.33){\makebox(0,0)[lc]{1}}
\put(131.67,42.33){\makebox(0,0)[rt]{2}}
\put(93.67,42.33){\makebox(0,0)[rt]{1}}
\put(150.67,32.67){\makebox(0,0)[lc]{1}}
\put(145.00,26.33){\makebox(0,0)[ct]{2}}
\end{picture}

\vspace{-1.0cm}

\noindent
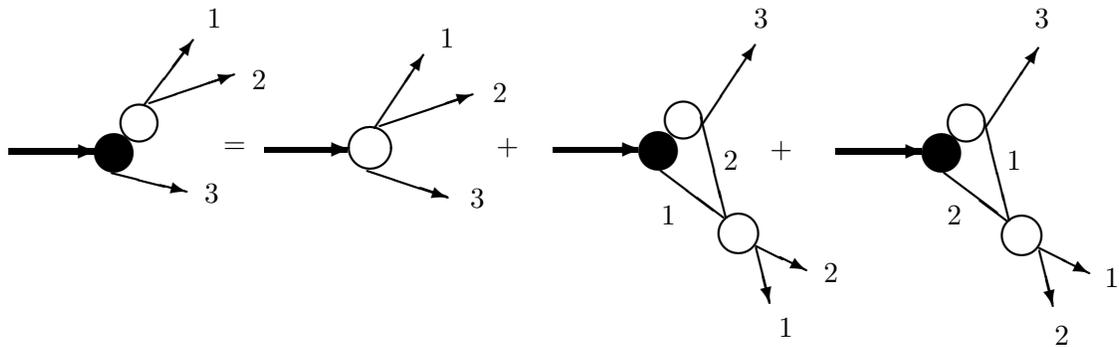
Fig.3. Graphic representation of the equation for the
amplitude $ A_1 (s,s_{12}) $ (formulae (4)-(6)).

\vspace*{4.0cm}

\unitlength 1.00mm
\linethickness{0.4pt}
\hspace{-1.2cm}
\begin{picture}(140.00,89.02)
\put(140.00,50.00){\vector(1,0){0.2}}
\put(20.00,50.00){\line(1,0){120.00}}
\put(70.00,80.00){\vector(0,1){0.2}}
\put(70.00,20.00){\line(0,1){60.00}}
\put(140.00,50.00){\vector(1,0){0.2}}
\put(109.67,50.00){\line(1,0){30.33}}
\put(140.00,50.00){\vector(1,0){0.2}}
{\thicklines\put(109.67,50.00){\vector(1,0){30.33}}}
\put(109.67,51.33){\line(0,-1){2.66}}
\put(30.00,50.00){\line(1,0){25.00}}
{\thicklines\put(30.00,50.00){\line(1,0){25.00}}}
\put(30.00,51.33){\line(0,-1){2.66}}
\put(55.00,51.33){\line(0,-1){2.66}}
{\thicklines \put(90.00,65.00){\line(0,-1){30.00}}}
\put(88.67,65.00){\line(1,0){2.66}}
\put(88.67,35.00){\line(1,0){2.66}}
\put(109.67,48.00){\makebox(0,0)[ct]{${(m_1 + m_3)}^2$}}
\put(92.00,65.00){\makebox(0,0)[lb]{$s_{13}^{+}$}}
\put(92.00,35.00){\makebox(0,0)[lt]{$s_{13}^{-}$}}
\put(88.67,55.33){\makebox(0,0)[rc]{1}}
\put(75.67,84.67){\circle{8.69}}
\put(75.67,84.67){\makebox(0,0)[cc]{$s_{13}$}}
\put(42.33,52.00){\makebox(0,0)[cb]{2}}
\put(30.00,48.00){\makebox(0,0)[ct]{$s_{13}^{-}$}}
\put(55.00,48.00){\makebox(0,0)[ct]{$s_{13}^{+}$}}
\end{picture}

\vspace*{-0.5cm}

\noindent
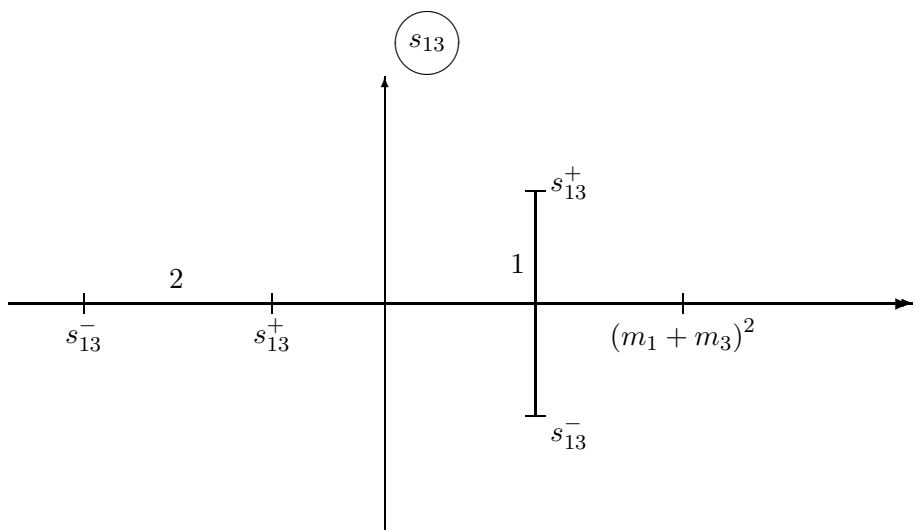
Fig.4. The contours of integration 1,2 in the complex plane $s_{13}$
for the function $ I_{J_1 , J_2}(s,s_0) $.

\end{document}